\documentclass[copyright,creativecommons]{eptcs}
\providecommand{\event}{Gabriel Ciobanu (Ed.): Membrane Computing and\\ Biologically Inspired Process Calculi 2009}
\providecommand{\volume}{15}
\providecommand{\anno}{2009}
\providecommand{\firstpage}{1}
\providecommand{\eid}{1}
\usepackage{breakurl}    

\usepackage{subfigure}
\usepackage{amssymb,amsbsy,latexsym,amsmath, euscript,amsfonts}
\usepackage{epic,eepic}
\usepackage[all]{xypic}
\usepackage[english]{babel}

\newcommand{\pair}[2] {( {#1},{#2} )}

\newcommand{\Rat} {\ensuremath{\mathbf{R}}}
\newcommand{\E} {\ensuremath{\mathbf{E}}}
\newcommand{\N} {\ensuremath{\mathbf{H}}}

\newcommand{\oi} {\sqsubseteq_I}
\newcommand{\ui} {\sqcup_I}

\newcommand{\aarrow}[3]{\xrightarrow[\hfill\circ]{#1,#2,#3}}

\newcommand{\vaa}[1]{{#1}^{\circ}} 

\newcommand{\splitnew}[1]{\bigtriangledown^{\sharp}(#1)}
\newcommand{\splitnewone}[1]{\bigtriangledown^{\sharp_1,\sharp_2}(#1)}

\newcommand{\f}{\ar@{-}} 

\newcommand{\lrtrans}[4]{#1{#2,#3,#4}}
\newcommand{\lrarrow}[3]{\lrtrans{\xrightarrow}{#1}{#2}{#3}}

\newcommand{\mul}[1] {[\![#1]\!]}

\newcommand{\funstyle}[1]{\ensuremath{\mathsf{#1}}}
\newcommand{\Paths}{\funstyle{Paths}}
\newcommand{\FPaths}{\funstyle{FPaths}}
\newcommand{\Distr}{\funstyle{Distr}}
\newcommand{\SDistr}{\funstyle{SDistr}}
\newcommand{\ADistr}{\funstyle{ADistr}}
\renewcommand{\Pr}{\ensuremath{\mathbf{P}}}

\newcommand{\Prm}{\ensuremath{\Pr^-}}
\newcommand{\Prp}{\ensuremath{\Pr^+}}
\newcommand{\Adv}{\funstyle{Adv}}
\newcommand{\Reach}{\funstyle{Reach}}
\newcommand{\flabel}{\ensuremath{\mathbf{L}}}

\newcommand{\lts}{\funstyle{LTS}}
\newcommand{\source}{\funstyle{source}}
\newcommand{\lab}{\funstyle{label}}
\newcommand{\rate}{\funstyle{rate}}
\newcommand{\target}{\funstyle{target}}

\newcommand{\abs}[1]{{#1}^{\circ}}

\newcommand{\grammarprod}{\ensuremath{\mathrel{\mathop:}\mathrel{\mathop:}=}}
\newlength{\grammarop}
\newcommand{\pro}{\mathop{\makebox[\grammarop]{$\grammarprod$}}}

\newcommand{\irule}[2]{\frac{\textstyle\rule[-1.3ex]{0cm}{3ex}#1}%
{\textstyle\rule[-.5ex]{0cm}{3ex}#2}}

\newtheorem{theorem}{Theorem}[section]

\newtheorem{definition}[theorem]{Definition}

\newtheorem{example}[theorem]{Example}

\title{Abstract Interpretation  for Probabilistic  Termination of Biological Systems}
\author{Roberta Gori
\institute{Dipartimento di Informatica}
\institute{Universita' di Pisa\\
Largo Pontecorvo 2 \\
Pisa, Italy}
\email{gori@di.unipi.it}
\and
Francesca Levi 
\institute{Dipartimento di Informatica}
\institute{Universita' di Pisa\\
Largo Pontecorvo 2 \\
Pisa, Italy}
\email{levifran@di.unipi.it}
}
\def\titlerunning{Abstract Interpretation  for Probabilistic  Termination of Biological Systems}
\def\authorrunning{Roberta Gori \& Francesca Levi}
\begin{document}
\maketitle

\begin{abstract}
In \cite{GL08}   the authors applied  the {\em Abstract Interpretation} approach for  approximating   the probabilistic semantics of  biological systems, modeled specifically using  the 
  Chemical Ground Form calculus \cite{Ca07}.  
The methodology is based on the idea of representing a set of experiments, which differ only for the initial concentrations, by abstracting
the multiplicity of reagents  present in a solution,  
using intervals. In this paper, we refine the approach in order to address {\em probabilistic termination properties}.
More in details, we introduce a refinement of the abstract LTS semantics and we abstract
the probabilistic semantics using a variant of {\em Interval Markov Chains} \cite{SVA06,FLW06,Hu05}.
The abstract probabilistic model
safely approximates a set of concrete experiments   and reports conservative \emph{lower} and  \emph{upper} bounds   for  probabilistic termination.
\end{abstract}




\section{Introduction}
\label{s:intro}
Process calculi, originally designed for modeling distributed and mobile systems, are nowadays one of the most popular formalisms for the specification of biological systems.  In this new application domain, a great effort has been devoted
for adapting 
  traditional models to characterize the molecular and biochemical aspects of biological systems.
Among them \cite{RPSCS04,PQ05,Ca04}, stochastic calculi, based on $\pi$-calculus
\cite{Pr05,PRSS01}, capture  the fundamental \emph{ quantitative} aspect
(both time and probability) of real life applications.
The use of a process calculus as a specification language  offers a range of well established methods for  analysis and  verification
that could now be applied to biological system models. These techniques can be applied to complex biological systems in order to test hypotheses and to guide future  \emph{ in vivo} experimentations. 
   Stochastic simulators, e.g. \cite{RSS01,PC04,PC07} for $\pi$-calculus, are able to realize \emph{ virtual experiments} on biological system models, while model checking techniques, recently extended also to 
 probabilistic and stochastic models \cite{HKN06,Kw03},  support the validation of  temporal properties.

However,  the  practical  application of automatic tools to biological systems revealed serious limitations.
One specific feature of biological processes is that 
they are 
composed by a huge number of processes with identical behavior, such as thousands of molecules
of the same type. 
Moreover,  typically  the exact concentrations of molecules are not  known, meaning that the hypotheses have  to be tested 
with respect to different  scenarios. Thus, different experiments have to be realized  and 
the state space of the models to be analyzed is often very large (even infinite). 

 \emph{Static analysis} techniques provide automatic and decidable methods for establishing
properties of programs, by computing safe approximations of the (run-time) behavior.
This approach has been successfully applied to {\em purely qualitative}  process calculi
for distributed and mobile systems, and recently also to biologically inspired process calculi, in order to validate safety as well as more complex  temporal properties \cite{BDNN01,Fe01,LM04,NNH02,GL05,GL06,NNP04}.

In \cite{GL08} we have proposed an approximation technique, based on
\emph{ Abstract Interpretation} \cite{CousotC77,CousotC79}, able
to address {\em probabilistic temporal properties} for a simple
 calculus,  the \emph{ Chemical Ground Form} (CGF)\cite{Ca07}.
CGF is a  fragment 
of stochastic $\pi$-calculus  which  is rich enough for 
modeling the dynamics of biochemical reactions.  The abstraction is based on the idea of
approximating the  information about the multiplicities of reagents,  present in a solution,  by means of
intervals of integers \cite{CousotC76}.
The approach computes an {\em abstract probabilistic semantics} for an abstract system, which  approximates the probabilistic semantics, namely 
the  {\em Discrete-Time Markov Chain} (DTMC), for any corresponding concrete system.
In particular, the validation of an  abstract system
 gives both \emph{lower} and  \emph{upper}  bounds  on the probability of temporal properties \cite{HJ94},
for a set of  concrete systems ({\em experiments}) differing only for the  concentrations of reagents.

{\tiny
\begin{figure}
\[
\def\objectstyle{\scriptstyle}
\def\labelstyle{\scriptstyle}
 \xymatrix{ 
 { LTS} \ar[rrrr]^{ \N}   \ar[d]^{\alpha_{lts}}    &  & & & { DTMC}  \ar[d]^{\alpha_{mc}}  \\
\abs{{ LTS}}  \ar[rrrr]^{ \vaa{\N}}& & & & \abs{{ IMC}} }
\]
\caption{The complete picture}\label{fig1000}
\end{figure}
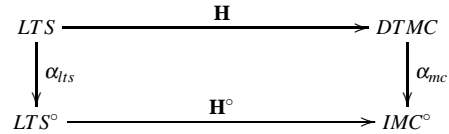}

The  methodology is illustrated in Fig. \ref{fig1000}. As usual,
the DTMC of a concrete system is derived   from the LTS semantics, by calculating the probability of each move.
The technique of abstraction is based on the definition of a suitable abstract LTS semantics for abstract systems, which support
the derivation of  an abstract probabilistic model, represented by an \emph{Interval Markov Chains} \cite{SVA06,FLW06,Hu05}. In 
Interval Markov Chains
transitions are labeled with intervals of probabilities, representing  the uncertainty about the concrete probabilities; consequently, the validation of temporal properties
reports  \emph{lower} and  \emph{upper}  bounds, rather than
exact values, which are obtained by considering the \emph{ worst-case}
and \emph{ best-case} scenario w.r.t. all   non-deterministic choices.
Obviously, the key step of the translation from abstract LTS into the Interval Markov Chain consists in the computation of intervals of probabilities from the information reported by abstract transition labels.
A quite precise approximation is achieved  because the information reported by transition labels is profitably
exploited in order
to capture also  relational information.

Unfortunately, 
if one is interested in proving 
more complex properties of biological systems, such as {\em probabilistic termination
}  \cite{ZC08}, 
the previously  proposed abstraction is not sufficiently powerful. 
For {\em probabilistic termination} we have to calculate the probability to reach a {\em terminated} state, e.g. a state where the probability to move in any other state is zero. 

{
\begin{figure}
\entrymodifiers={++[o][F-]}
\UseComputerModernTips
\[\xymatrix @C+2pc {
X \ar@(ul,dl)[]_{!a}\ar@(u,u)[r]^{?b}   & Y\ar@(dr,ur)[]_{!b}\ar@(d,d)[l]^{?a} }\]
\caption{Groupie automata}\label{f:groupies}
\end{figure}
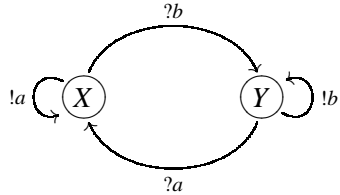}

To illustrate  probabilistic termination, we consider  the  ''groupies'' example proposed by Cardelli in several tutorials  on biochemistry and also reported in   \cite{Ca09}.
The idea is to study how a set of  entities collectives behave.
The behavior of  a single entity is represented by the automaton in Fig. \ref{f:groupies}; it  has two possible states, $X$ and $Y$. A single automaton performs no interaction, while it may interact with other automata. Two automata in state $X$ are stable since they both offer $!a$ and $?b$ and no interaction are possible. Analogously for two automata in state $Y$. If one automata is in state $X$ and another is in state $Y$ then either they can interact on channel $a$ and both move to state $X$ or  they can interact on channel $b$ and both move to state $Y$.
No matter how many   automata are in state $X$ or  in state $Y$ initially,
   eventually the groupies form a single homogeneous population of all  $X$  or of all $Y$. 
Thus, these systems always   terminate, namely  they {\em universally} terminate.

The limitation of the   abstract LTS semantics, defined in  \cite{GL08},  is represented by {\em hybrid}  states, namely abstract states representing  concrete  {\em terminated} as well as {non terminated} states. It should be clear that, given an abstract state,  the {\em most precise} and {\em correct} intervals of probabilities, could be derived by considering the { \em minimum}
and {\em maximum} exact probabilities, for each concrete move, respectively.
Thus, for an hybrid state
we would obtain very  approximated  intervals of probabilities, such as $[0,1]$, both for the self-loop and for any other move. 
This information says that some concrete states may loop forever, while others may move somewhere else.
As a consequence, the lower and upper bound probabilities to reach a terminated state, from an hybrid state,  are typically
zero and one, respectively.
This is the case of example for the CGF specification of  groupies example, previously commented.

In order to better capture probabilistic termination, we propose in this paper a refinement of our approach, based on a modification of the abstract LTS semantics. More in details, the abstract transition relation is refined so that 
terminated and non-terminated states are properly separated, and consequently hybrid states are never  generated.
To this aim, it may be necessary to replace a
 single abstract transition, corresponding to a given reaction,  by a set of abstract transitions, leading to different abstract target states. Such  distinct abstract transitions model  the same reaction but with different concentrations of reactants. This situation induces a notion of {\em conflict} between abstract transitions; indeed, the corresponding reaction, for each concrete state, is approximated by exactly one of those abstract transitions.
In this context, the labels of transitions precisely identify the  interaction
and can naturally  be exploited to capture conflicts between abstract transitions.

Once the abstract LTS semantics has been refined, the remaining  problem is  to
generalize the translation from the abstract LTS  to the abstract probabilistic model. In order to maintain  the information about conflict, recorded by abstract transition labels, we adopt a generalization of the original model, called
{\em Labeled Interval Markov Chains} (IMC). In IMC  the labels  permit to more accurately represent the set of distributions represented by the interval of probabilities. 
We show that the technique of \cite{GL08} for computing   intervals of probabilities from abstract
transition labels  can be successfully generalized,
by finding out a good trade-off between precision and 
complexity. 
Finally, the soundness of the proposed technique is formalized following the approach of \cite{GL08} (see also
\cite{DGG94,DJJL01,DJJL02,SVA06,Hu05}) which exploits suitable approximation orders, both on  abstract LTS and
on  IMC.

The paper is organized as follows. Section \ref{s:cgf} introduces the CGF calculus and the LTS semantics, while 
 Section \ref{s:mc} shows the  probabilistic semantics in terms of a  DTMC. 
Section \ref{s:ltsa} presents the refined abstract LTS semantics. 
  Section \ref{s:markova} introduces the IMC model and finally,
 Section \ref{s:aps} presents the effective method to derive  the abstract probabilistic semantics.

 \section{Chemical Ground Form}
\label{s:cgf}

The CGF calculus \cite{Ca07} is a fragment of stochastic $\pi$-calculus \cite{Pr05,PC04}
without communication. Basic actions are related to {\em rates}, which are
the parameters of the exponential distribution.
We present the {\em labeled transition system} (LTS) semantics of CGF, proposed in \cite{GL08}, which supports more precise abstractions
with respect to the original proposal of \cite{Ca07}.
In this approach, 
processes are labeled,  and transitions
record information about
the labels of the actions which participate to the move, about their  rates, and about their number of occurrences (in place of  the rate of the move as in \cite{Ca07}). 

The syntax of (labeled) CGF is defined in
Table \ref{tab:syn-cgf}. We  consider a set $\cal N$ (ranged over by
$a,b,c, \ldots$)  of {\em names}, a set $\cal L$ (ranged over by
$\lambda, \mu \ldots$)  of {\em labels}, and a set $\cal X$ (ranged over by $X$,$Y$,....)
of {\em variables} (representing reagents).

\begin{table} 

  \hrule
  \[\begin{array}{llll}
    E &\pro 0 \mid  X = S, E & & \mbox{Environment} \\
    S &\pro 0 \mid \pi^\lambda.P + S & & \mbox{Molecules}\\
    P &\pro 0 \mid X|P & & \mbox{Solutions}\\
    \pi & \pro a_r \mid \bar a_r \mid \tau_r \ \ \ r \in \mathbb{R}^+& & \mbox{Basic Actions}
  \end{array}\]
\hrule
  \caption{Syntax of CGF}
   \label{tab:syn-cgf}
   \end{table}

A CGF is defined as a pair $ (E,P)$ where $E$ is a \emph{species environment}
and $P$ is a {\em solution} .
The environment $E$ is a (finite) list of  reagent definitions 
$X_i = S_i$ for  distinct variables $X_i$ and molecules $S_i$.
We assume that the environment $E$  defines all the  reagents  of solution  $E$.
A \emph{molecule} $S$ 
may do nothing, or may change after a delay or may
interact with other reagents. A standard notation is adopted: $\tau_r$ represents a delay at rate $r$; 
$a_r$ and  $\bar a_r$ model, respectively, the input and output on channel $a$ at rate $r$. A solution $P$ is a parallel composition
of variables, that is
a finite list of reagents.

Labels are exploited in order to distinguish the actions which participate to a move.
To this aim, we consider CGF $(E,P)$, where $E$
is {\em well-labeled}, meaning that the labels of basic actions
are all distinct. 
Moreover, given
a label $\lambda \in {\cal L}$, we
use the notation $E.X.\lambda$ to indicate the process $\pi^\lambda.P$ 
provided that $X = \ldots +
\pi^\lambda.P + \ldots$ is the definition of $X$ occurring in $E$.  
We may also use ${\cal L}(E.X)$ for the set of labels
appearing in the definition of $X$  in $E$.

 The  semantics is based on the natural representation
of solutions as multisets of reagents.
A {\em multiset} is a function $M:{\cal X}  \rightarrow \mathbb{N}$.
In the following, we use ${\cal M}$ for the set of multisets and we use
$\mul{P}$ for the multiset of reagents corresponding to a solution $P$.
Moreover,
we call $M(X)$ the multiplicity of reagent $X$ in the multiset $M$.  We may also represent multisets as sets  of pair $(m, X) $, where $m$ is the multiplicity of reagent $X$, using a standard notation, where
the pairs with multiplicity $0$ are omitted.
Over multisets we use the standard operations of sum and difference  $\oplus$ and $\ominus$, such that
$\forall X\in {\cal X}$: $M\oplus N(X)=M(X)+N(X)$ and $M\ominus N(X)=M(X) \widehat{-} N(X)$ where
$n \widehat{-} m=n-m \mbox{  if }n-m \geq0, \;0\mbox{  otherwise.}$


The evolution  of a solution (w.r.t. a given environment $E$) is described by a  labeled
transition relation of the form\\ 
\hspace*{7cm}$M \lrarrow{\Theta}{\Delta}{r} M'$\\
where  $r \in \mathbb{R}^+$  is a rate, $\Theta \in \widehat{{\cal L}}= {\cal L}   \cup ({\cal L}  \times {\cal L})$, $\Delta \in \widehat Q=\mathbb{N}\cup (\mathbb{N} \times \mathbb{N} )$ such that
$arity(\Theta)=arity(\Delta)$.
Here, $\Theta$ reports the label (the labels) of the basic action (the basic actions), which participate
to the move,  $\Delta$ reports consistent information about the
multiplicity, and $r$ is the related rate.



The transition relation  for multisets is defined by the rules
Table~\ref{tab:tran-cgf} (we are tacitly assuming to reason w.r.t. a given  environment $E$).  
Rule ({\bf Delay}) models the
move of a process ${\tau_r}^{\lambda}.Q$ appearing in the definition of a reagent $X$. The transition records  
the  label $\lambda$ together with the multiplicity of $X$ (e.g 
$M(X)$) as well as the rate  $r$.
Rule ({\bf Sync}) models the
synchronization between two complementary processes 
${a_r}^{\lambda}.Q_1$ and 
${\bar {a_r}}^{\mu}.Q_2$ appearing 
in the definition reagents $X$ and $Y$ (that may even coincide).
The transition records  
the labels $\lambda$ and $\mu$ together with the multiplicities of $X$ and $Y$ (e.g 
$M(X)$ and $M(Y)$) as well as the rate  $r$.


\begin{table} 

  \hrule
\[\begin{array}{ll}
(\mbox{{\bf Delay}})\;\;\;\;
\irule{E.X.\lambda = {\tau_r}^{\lambda}.Q }
{{M}\lrarrow{\lambda}{ {M}(X)}{r} 
    (M \ominus (1,X)) \oplus \mul{Q}
}
\vspace{0.3cm}\\
(\mbox{{\bf Sync}})\;\;\;\;
\irule{E.X.\lambda = {a_r}^{\lambda}.Q_1  \qquad
        E.Y.\mu = {\bar {a_r}}^{\mu}.Q_2
}
{ 
{M} \lrarrow{(\lambda,\mu)}{ \pair {{M}(X)} {{M}(Y)}
}{r} ((M \ominus
      (1,X)) \ominus (1,Y)) \oplus \mul{Q_1} \oplus \mul{Q_2} 
}
\end{array}  \] 
 \hrule 
  \caption{Transition relation} \label{tab:tran-cgf} 
\end{table}

We denote with $\lts((E, M_{0}))=(S,\rightarrow,M_{0},E)$ the LTS, obtained 
 as usual by transitive closure, starting from the initial state $M_0 \in S$, w.r.t. to environment $E$. 
Note that, since environments are well-labeled, e.g. basic actions have distinct labels, the transitions from a state of the LTS are decorated by distinct labels
too.  Moreover, we use ${\cal LTS}$ to denote the set of LTS.

%


In the following, given a transition $t=M \lrarrow {\Theta}{\Delta} r M'$ we use
$\lab(t)$ to denote its label $\Theta$, and $\source(t), \target(t)$ to
denote its source state $M$ and target $M'$, respectively. Similarly, for a set of transitions
$TS$, we use $\lab(TS)=\bigcup_{t \in TS}
\lab(t)$.
We also use $
\mathsf{Ts}(M,M') = \{ t \mid \source(t)=M \ \mbox{and} \ \target(t)=M' \}$ and $
\mathsf{Ts}(M) = \{ t \mid \source(t)=M  \}$
for describing the transitions from a multiset $M$ to a multiset $M'$, and all  transitions leaving from multiset $M$, respectively.








\section{Probabilistic Semantics}
\label{s:mc}
We introduce the  probabilistic model of DTMC and we briefly discuss the notion of {\em probabilistic termination} \cite{ZC08}.
We also introduce the probabilistic semantics of CGF proposed in \cite{GL08}.
 \vspace*{0.3cm}
\\
{\bf Dicrete-Time Markov Chains.} 
\vspace*{0.3cm}
\\ 
Given a finite or countable set of states $S \subseteq {\cal M}$ we denote with

\[\begin{array}{llll}
 \SDistr(S) = \{ \rho \mid \rho \colon S \to [0,1] \}, 
&\;  \Distr(S) = \{ \rho \mid \rho \in \SDistr(S)  \text{ and }
  \sum_{M \in S} \rho(M) = 1 \}&
\end{array}\]

the set of (discrete) probability  \emph{pseudo-distributions} and of \emph{distributions}  on $S$,
respectively.
\begin{definition}[DTMC]
  A DTMC is a tuple $(S, 
  \Pr, \flabel, M_0)$ where: (i) $S \subseteq {\cal M} $ is a {\em finite or countable} set of states, $M_0 \in S$ is the {initial state};  (ii)   $\Pr \colon S \to \Distr(S)$ is the 
   \emph{probability transition   function};
 (iii) $\flabel: S  \to (S \to  \wp ({\widehat{{\cal L}}}))$ is  a {\em labeling function}.
\end{definition}
In DTMC state transitions are equipped with probabilities, e.g. 
$\Pr(M)(M')$ reports the probability of moving from state $M$ to state $M'$.
In addition, 
 $\flabel(M)(M')$ reports the set of labels corresponding to the moves  from state $M$ to state $M'$.
Notice  that we adopt a labeled version of the model in order to simplify the correspondence with the abstract models; the labels do not modify the probability distributions in the concrete model.  We use ${\cal MC}$ for the set of DTMC.

We are interested in {probabilistic termination}, e.g. on the probability to reach  a  state, which is {\em terminated}. Given a DTMC $ (S,  \Pr, \flabel, M_0)$, we say that a state $M\in S$  is 
 \emph{terminated}  iff  $\Pr(M)(M')=0$, for each $M' \in S$ with $M' \neq M$.

The probability to reach a terminated state can be formalized
by associating a probability measure to  paths
of a DTMC.
Let $ (S,  \Pr,\flabel, M_0)$ be a DTMC.
A 
  \emph{path} $\pi$ is a non-empty  sequence of states of $S$.  We
  denote the $i$-th state in a path $\pi$  by $\pi[i]$, and the
  length of $\pi$ by $|\pi|$.  The
  set of  (resp. finite) paths over $S$ is denoted by (resp. $\FPaths(S)$) $\Paths(S)$, while
$C(M)$ denotes the set of paths starting from the state $M \in S$.
In the following, for $M \in S$ and $\Pi \in C(M)$, 
$\Pr_M(\Pi)$ stands for the probability of the sets of paths $\Pi$
(see  \cite{KSK76} for the standard definition).



\begin{definition}[Probabilistic Termination]
  Let $ mc=(S,  \Pr,\flabel, M_0)$ be a DTMC.  The probability of
  reaching a {\em terminated  state},  from   $M \in S$, is $
    \Reach_{mc}(M) = \Pr_M(\{ \pi \in C(M) \mid \pi[\,  |\pi|\, ]   \text{ is terminated, and }  \forall j, 0\leq j\leq |\pi|, \pi[j] \text{ is non-terminated} \}).$
\end{definition}
\vspace*{0.3cm}
{\bf Derivation of the DTMC.} \vspace*{0.3cm} \\
The derivation of a DTMC from the LTS is based on the computation
of the   probability of moving from $M$ to $M'$, for any  $M$ and $M'$. 
To this aim, we
 extract the rate
corresponding to the move from $M$
to $M'$  by exploiting the information reported by transition labels.

Formally, for a transition
$t=M \lrarrow {\Theta} {\Delta}r M'$ we define the corresponding {\em rate} as follows,

\[\rate(t)= \left \{ \begin{array}{llll} n \cdot r & \hspace*{0.3cm}
\Theta= \lambda,\Delta=n,\\
n \cdot (m \widehat{-}1) \cdot r & \hspace*{0.3cm}
\Theta=(\lambda,\mu), \Delta= (n,m),
 \lambda,\mu \in {\cal L}(E.X),\\
n \cdot m \cdot r & \hspace*{0.3cm}
\Theta=(\lambda,\mu), \Delta= (n,m),   \lambda \in {\cal L}(E.X), \mu \in {\cal L}(E.Y), X \neq Y. 
\end{array}\right.
\]

As usual, for computing $\rate(t)$ it is necessary to take into 
account the number of distinct transitions $t$
 that may occur in the multiset $M$.
Thus, 
the rate $r$ of the basic action (actions) related to  $\Theta$ is multiplied by the number of distinct combinations appearing in $M$ (by exploiting the information recorded by $\Delta$).

Then, we introduce functions $\Rat: S \times S \to \mathbb{R}^{>=0}$
and $\E: S \to \mathbb{R}^{>=0}$, such that
\[ \begin{array}{l}
\Rat(M,M')=\sum_{t \in \mathsf{Ts}(M,M')} \rate(t)\ \ \
\E(M)=  \sum_{M' \in S} \Rat(M,M').
\end{array}\]

Intuitively, $\Rat(M,M')$ reports the rate corresponding to the move from $M$
to $M'$, while $ \E(M)$ is the {\em exit rate}. 
Finally, the probability of moving from $M$ to $M'$
is computed from  $\Rat(M,M')$ and from the exit rate $\E(M)$, in a standard way.

\begin{definition}
\label{d:normal}
 We define a {\em probabilistic translation} function $\N :{\cal LTS} \rightarrow  {\cal MC}$
such that 
    $\N((S,\rightarrow,M_{0},E)) = (S,\Pr,\flabel,M_0)$, where 
\begin{enumerate}
\item
    $\Pr : S \to \Distr(S)$ is the \emph{probability transition
      function}, such that for each $M,M' \in S$:
      \begin{description}
\item[a)]
if $\E(M) >0$, then  $\Pr(M)(M')= \Rat(M,M')/ \E(M)$; 
\item[b)]
 if $\E(M) =0$, then $\Pr(M)(M)= 1$,  and  $\Pr(M)(M')= 0$ for $M'\not=M$. 
\end{description}
\item $\flabel: S  \to (S \to  \wp ({\widehat{{\cal L}}}))   $ is  a {\em labeling function}, such that, for each
 $ M,M' \in S$, $\flabel(M,M')=\lab(\{t \in \mathsf{Ts}(M,M') \mid \rate(t)> 0\}).$
\end{enumerate}
\end{definition}


Due to the particular labeling of the LTS semantics, also 
 the DTMC, modeling the probabilistic semantics of a CGF process, satisfies the properties that all transitions  leaving from a state, are decorated by distinct labels.


\begin{example}\label{ese2} 
The example of groupies, commented in the Introduction, can be formalized by the following environment,
$E \pro \ \   X=a_{r}^{\lambda}.X+ \bar{b}_{r}^{\delta}.Y, 
 Y= \bar{a}_{r}^{\mu}.X + b_{r}^{\eta}.Y$.

Reagents  $X$ and $Y$ may interact together  in two possible ways;  {\em either}  along channel
$a$  {\em or}  along channel
$b$; both reactions have  the same rate $r$. The former case models  a {\em duplication}
of $X$, while the latter case models a {\em duplication}
of $Y$.

Fig. \ref{fig300} illustrates the LTS  and the corresponding DTMC,
for the CGF $(E,M_0)$, 
where  \\
 $\begin{array}{llll}
M_{0}=\{(1,X), (2,Y)\} \ \ 
M_{1}=\{(2,X), (1,Y)\} \ \  
M_{2}=\{(3,X)\}\ \
M_{3}=\{(3,Y)\} \ \   
\end{array}$

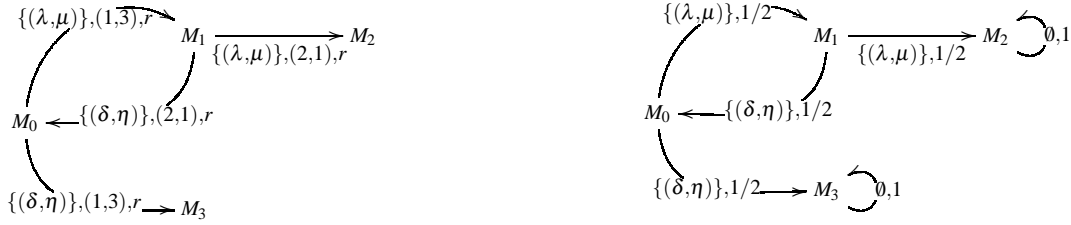
\begin{figure}
\[\begin{array}{ll}
\def\objectstyle{\scriptstyle}
\def\labelstyle{\scriptstyle}
 \xymatrix@C+2pc@R-0.3pc{ 
  & M_{1} \ar[r]_{\{(\lambda, \mu)\}, (2,1),r}  
  \ar@(d,r)[dl]|{\{(\delta, \eta)\}, (2,1),r}    &   M_{2}   \\
M_{0} \ar@(u,ul)[ur]|{\{(\lambda, \mu)\}, (1,3),r}   \ar@(d,l)[dr]|{\{(\delta, \eta)\}, (1,3),r}   \\
& M_{3} 
   }
& \hspace*{3cm}
\def\objectstyle{\scriptstyle}
\def\labelstyle{\scriptstyle}
\xymatrix@C+2pc@R-0.6pc{ 
 & M_{1} \ar[r]_{\{(\lambda, \mu)\}, 1/2}  
  \ar@(d,r)[dl]|{\{(\delta, \eta)\}, 1/2}    &   M_{2}   \ar@(dr,ur)[]|{\hspace* {0.3cm}\emptyset, 1}
\\
M_{0} \ar@(u,ul)[ur]|{\{(\lambda, \mu)\}, 1/2}   \ar@(d,l)[dr]|{\{(\delta, \eta)\}, 1/2}   \\
& M_{3} 
\ar@(dr,ur)[]|{\hspace* {0.3cm}\emptyset, 1}
  }
\end{array}\]
\caption{The LTS and the corresponding DTMC}\label{fig300}\end{figure}

The LTS reports for each state, except for states $M_2$ and $M_3$, two transitions:
 label $(\lambda, \mu)$ models the duplication of $X$, while label $(\delta, \eta)$ models
the duplication of $Y$. The transitions record also the multiplicities of reagents $X$ and $Y$
and the corresponding rate.
As a consequence, in the DTMC, states $M_2$ and $M_3$ are terminated. By contrast, the states
$M_0$ and $M_1$ have two different moves with the same probability.
By calculating the probability to reach a terminated state from $M_0$ we obtain exactly 1.
Indeed,  the probability to be stuck in the  loop  $M_0$-$M_1$  is zero.
\hfill$\Box$
\end{example}



\section{Abstract LTS}
\label{s:ltsa}
The abstract LTS semantics uses the same abstraction of multisets of \cite{GL08}, based on the approximation of the multiplicity
of reagents by means of intervals of integers \cite{CousotC76}. Instead, 
 the abstract transition relation is refined, and the related notions, needed for expressing soundness, are adapted accordingly.
\vspace*{0.3cm}\\
{\bf Abstraction of states.} \vspace*{0.3cm} \\
 We 
 adopt intervals of integers, ${\cal{I}}=  \{[m,n] \mid  m\in \mathbb{N},n \in \mathbb{N} \cup \{\infty\} \wedge
m\leq n \}.$
Over intervals we consider the standard order $\oi$, such that
$I \oi J$ iff $\tt{min}(I),\tt{max}(I) \in J$. Moreover, we use 
$\ui$ for the corresponding l.u.b..

The abstract states are defined 
 by replacing
 multiplicities   with  intervals of multiplicities.
Therefore, 
an {\em abstract state} is a function $\vaa{M}:{\cal X}  \rightarrow \cal{I}$.
We also use $\vaa{\cal M}$ for the set of abstract states.

Obviously, given a multiset $M$, there exists an abstract multiset $\vaa{M}$, which is its { most precise} approximation. Indeed, 
each multiplicity, such as $n$, can be replaced with the  exact interval $[n,n]$; for simplicity, we may even
use $n$ as a shorthand of $[n,n]$.
In the following, $\alpha(M)$ stands for the
{\em best abstraction} of a multiset $M$.
Moreover, we use $\vaa{M} [I/X]$ for denoting the abstract state where the abstract multiplicity of reagent $X$ is replaced by the interval $I \in {\cal{I}}$.
We adopt abstract operations of sum and difference, such that $\forall X\in {\cal X}$,
\[\begin{array}{l}
 \vaa{M}\vaa{\oplus} \vaa{N}(X)= \vaa{M}(X)+\vaa{N}(X),\ \ \
 I + J  =
[ min(I) + min(J), max(I) + max(J) ] \\
\vaa{M}\vaa{\ominus} \vaa{N}(X)=\vaa{M}(X)-\vaa{N}(X),\ \ \
I - J  =
[ min(I) \widehat{-} max(J), max(I) \widehat{-} min(J) ]
\end{array}\]
It  is immediate to define the following approximation order over abstract states.

 \begin{definition}[Order on States]
\label{d:nuovo}
Let $\vaa{M}_{1}, \vaa{M}_{2} \in \vaa{\cal
M}$, 
we say that $\vaa{M}_1 \vaa{\sqsubseteq} \vaa{M}_2$ iff, for each reagent $X\in {\cal X}$,
$\vaa{M}_{1}(X) \oi      \vaa{M}_{2}(X)$.
\end{definition}

The relation between multisets and abstract states is formalized
as a Galois connection \cite{CousotC79}.
The {\em abstraction function} $\alpha:{\cal P}({\cal M}) \rightarrow
\vaa{\cal M}$ reports the {\em best approximation}
for each set of multisets $S$; the  l.u.b. (denoted by $\vaa{\sqcup}$)
of  the best abstraction of each  $M \in S$.
Its counterpart is the 
{\em concretization function} $\gamma:\vaa{\cal M} \rightarrow {\cal P}({\cal M})$
which reports the set of multisets represented by
an abstract state. We refer the reader to 
\cite{GL08} for the properties of  functions $(\alpha,\gamma)$.

\begin{definition}
 \label{d:gc}
We define $\alpha:{\cal P}({\cal M}) \rightarrow
\vaa{\cal M}$  and $\gamma:\vaa{\cal M} \rightarrow {\cal P}({\cal M})$ such that, for each $S \in {\cal P}({\cal M}) $ and $\vaa{M} \in 
\vaa{\cal M}$: (i) $\alpha(S)= \vaa{\bigsqcup}_{M \in S}\alpha(M)$; (ii) $\gamma(\vaa{M})=\{ M' \mid \alpha(M') \vaa{\sqsubseteq} \vaa{M}$\}.
\end{definition}

 %
{\bf Abstract transitions.}\vspace*{0.3cm}\\
The semantics of  \cite{GL08} uses abstract transitions of the form $
\vaa{M}_1 \aarrow \Theta {\vaa{\Delta}} {r} {\vaa{M}_2} $
where $\Theta \in \widehat{{\cal L}}$,  $\vaa{\Delta} \in \vaa{\widehat{Q}}=  {\cal I}   \cup  ({\cal I} 
 \times {\cal I})$, with $arity(\Theta)=arity(\vaa{\Delta})$.
Similarly as in the concrete case, $\Theta$ reports the label (the labels) of the basic action (actions),
$\vaa{\Delta}$ reports consistent information about the {\em possible multiplicities}, while  $r$ is the  rate.

In the proposed approach, such a  transition is intended to approximate {\em all} the concrete moves, corresponding to  label $\Theta$, for each multiset $M_1$ approximated by the abstract state
$\vaa{M}_1$. This means that  there exists a concrete  transition ${M}_1 \lrarrow \Theta {{\Delta}} {r} {M}_2$, where the multiplicity (multiplicities)
$\Delta$ is included in the interval (intervals) $\vaa{\Delta}$, and  $M_2$ is approximated by the abstract state
$\vaa{M}_2$.

Let us consider the environment $E$ commented in Example \ref{ese2}
and a very simple abstract state such as $\vaa{M}_0=\{ ([1,2],X), ([1,2],Y)\}$. The abstract state $\vaa{M}_0$ describes a set of experiments; thus, the abstract  semantics has to model the system described by $E$, w.r.t. different initial concentrations.
For approximating
the {\em duplication} of $X$, i.e. 
the synchronization between   $X$ and $Y$ along channel $a$,
we would obtain

\[\vaa{M}_0 \aarrow  {(\lambda, \mu)} {([1,2],[1,2])}{r} {\vaa{M}_0}'\;
\mbox{    with ${\vaa{M}_0}' =\{ ([2,3],X), ([0,1],Y)\}.$}
\]  

In this way, however, a {\em hybrid} state ${\vaa{M}_0}' $ is introduced.
Actually, ${\vaa{M}_0}' $  represents terminated multisets, where the concentration of reagent 
$Y$ is zero, as well as non terminated multisets, where reagent  $Y$ is still available. 

It should be clear that 
the moves corresponding to $(\lambda,\mu)$ could be better approximated by adopting two different abstract transitions,

\[\begin{array}{ll}
\vaa{M}_0 \aarrow  {(\lambda, \mu)} {([1,2],[2,2])}{r} \vaa{M}_{1} \ \ (a)&\hspace*{1cm}
\vaa{M}_0 \aarrow  {(\lambda, \mu)} {([1,2],[1,1])}{r} \vaa{M}_{3}  \ \ (b)
\end{array}\]

where $\vaa{M}_{1}=\{ ([2,3],X), ([1,1],Y)\}$ and $\vaa{M}_{3}=\{ ([2,3],X), ([0,0],Y)\}$.
In this representation the labels capture a relevant information because they express a {\em conflict}. Actually, each multiset represented by $\vaa{M}_0$,
realizes a move corresponding to $(\lambda,\mu)$ which is abstracted {\em either} by  transition (a)  {\em or} by transition (b).

Table~\ref{tab:atran} presents the {\em refined abstract transition rules} 
(as usual, w.r.t.  a given  environment $E$).
The rules are derived from the concrete ones, by replacing multiplicities
with intervals of multiplicities. The following operators are applied both to the target state
and to the intervals, appearing in the transition labels, in order to properly split the intervals, such as $[0,n]$.

 For $X \in {\cal X}$, we define $\aleph(X)= \{(X=0), (X>0) \}$. Then, given
an abstract state $\vaa{M} \in {\cal M}$ and $\sharp \in \aleph(X)$ 
we define

\[\splitnew{ \vaa{M}}= \left \{ \begin{array}{llll}
         \vaa{M}[[0,0]/X]       & \mbox{if} \ \ \sharp=(X=0), \vaa{M}(X)=[0,n], n>0\\
\vaa{M}[[1,n]/X]       & \mbox{if} \ \ \sharp=(X>0), \vaa{M}(X)=[0,n], n>0\\
\vaa{M} & \mbox{otherwise}
\end{array} \right.
\]

With an abuse of notation, we may write $\splitnewone{\vaa{M}}$
in place of $\bigtriangledown^{\sharp_1}(\bigtriangledown^{\sharp_2}(\vaa{M}))$.
Similarly, for an interval $I=[n,m] \in {\cal I}$ and  $\sharp \in \aleph(X)$, 
\[I^{\sharp}= \left \{ \begin{array}{llll}
   [n,1]           & \mbox{if} \ \ \sharp=(X=0),  n \leq 1,\\
\mbox{$[2,m]$}     & \mbox{if} \ \ \sharp=(X>0),  n \leq 1, m \geq 2,\\
I & \mbox{otherwise}.
\end{array} \right.
\]


In the following we use $\vaa{{\cal LTS}}$ to denote the set of abstract LTS. We also assume that all  notations defined
for LTS are adapted in the obvious way.
Hence, we write
$\vaa{\lts}((E,\vaa{M}_{0}))=(\vaa{S},\rightarrow_{\circ}, $ $\vaa{M}_{0},E)$ for the abstract LTS,
obtained for the initial  abstract state $\vaa{M}_0$ by transitive closure.
 
\begin{table} 
  \hrule
\[\begin{array}{ll}
(\mbox{{\bf Delay-a}})\;\;\;\;  
\irule{E.X.\lambda = {\tau_r}^{\lambda}.Q \ \ \ \ \sharp \in \aleph(X)
}
{
{\vaa{M}     \aarrow{\lambda}{ (\vaa{M}(X))^{\sharp}}{r} 
    \splitnew{(\vaa{M} \vaa{\ominus}    \{(1,X)\}) \vaa{\oplus} \alpha(\mul{Q})}}}
\vspace{0.3cm}\\   (\mbox{{\bf Sync-a}})\;\;\;\;
  \irule{
                 E.X.\lambda = {a_r}^{\lambda}.Q_1  \qquad
        E.Y.\mu = {\bar {a_r}}^{\mu}.Q_2 \ \ \ \ \sharp_1 \in \aleph(X) \ \ \ \ \sharp_2 \in \aleph(Y)
      }
{
 \vaa{M} \aarrow {(\lambda,\mu)}
{ \pair {(\vaa{M}(X))^{\sharp_1}} {(\vaa{M}(Y))^{\sharp_2}}}
{r} 
  \splitnewone{ ((\vaa{M} \vaa{\ominus} \{(1,X)\}  ) \vaa{\ominus} \{(1,Y)  \} )
         \vaa{\oplus}\alpha(\mul{Q_1}) \vaa{\oplus}  \alpha(\mul{Q_2})}}    
  \end{array}  \] 
 \hrule 
\caption{Abstract transition relation}
  \label{tab:atran}
\end{table}

For the sake of simplicity  we have presented an  approximation where the number of states may be infinite. Further approximations  can be easily derived by means of widening operators (see \cite{GL08}).\vspace*{0.3cm}\\
{\bf Soundness.}\vspace*{0.3cm}\\
In the style of \cite{GL08}, we introduce
 an  approximation order $\vaa{\sqsubseteq}_{lts} $ over abstract LTS. In this way,
we can say that an abstract LTS $\vaa{lts}$  is a {\em sound  approximation} of a LTS $lts$ provided that  $\alpha_{lts}(lts)
\vaa{\sqsubseteq}_{lts} \vaa{lts}$; as usual, $\alpha_{lts}(lts)$ is the {\em best approximation}
of $lts$.


 \begin{definition}[Best Abstraction of LTS] 
\label{d:ltsba}
  We define $\alpha_{lts}: {\cal LTS} \rightarrow \vaa{{\cal LTS}}$, such that
$\alpha_{lts}((S,\rightarrow, M_0,E))$\\ = $(\{ \alpha(M)\}_{M\in S},\alpha({\rightarrow}), \alpha( M_0),E)$
where
$\alpha({\rightarrow})=\{\alpha(M)  \aarrow {\Theta} {\vaa{\Delta}} { r}  \alpha({M}_{1}) \mid {M}  \lrarrow {\Theta} {{\Delta}} { r}  {M}_{1}
\in \rightarrow\}$ and $\vaa{\Delta}$ is the best abstraction
of $\Delta$, derived component-wise.

\end{definition}

In the following, we assume to extend the order $\oi$ over intervals to
pairs of intervals; $\vaa{\Delta_1} \oi \vaa{\Delta_2} $ is defined component-wise.

\begin{definition}[Order on abstract LTS]\label{d:orderalts}
  Let $\vaa{lts}_i =  (\vaa{S}_{i},\rightarrow^{i}_{\circ}, \vaa{M}_{0,i},
E) $ with $i \in \{1,2\}$ be abstract LTS.
  For $\vaa{M}_{1}\in \vaa{S}_1, \vaa{M}_{2}\in \vaa{S}_2$, we say that $\vaa{M}_{1} \preccurlyeq_{lts} \vaa{M}_{2}$  iff exists a relation $R\subseteq \vaa{S}_1\times  \vaa{S}_2$ such that if $\vaa{M}_{1} R \vaa{M}_{2}$ then: (i) $\vaa{M}_{1} \vaa{\sqsubseteq} \vaa{M}_{2}$; and  (ii) there exists a surjective function   $H_{t}:\mathsf{Ts}(\vaa{M_{1}}) \rightarrow
  \mathsf{Ts}(\vaa{M_{2}})$ such that,
      for each
      $\vaa{t}_{1}\in \mathsf{Ts}(\vaa{M_{1}}),  $
     $\vaa{t}_{1} =\vaa{M}_{1} \aarrow{\Theta}{\vaa{\Delta}_{1}}{r} \vaa{N}_1$, $H_{t}(\vaa{t}_{1})=\vaa{t}_{2}$ where
   $\vaa{t}_{2}=  \vaa{M}_{2} \aarrow{\Theta}{\vaa{\Delta}_{2}} {r} \vaa{N}_2$,
  $\vaa{\Delta}_{1}\oi\vaa{\Delta}_{2}$
 and 
       $\vaa{N}_{1}R \vaa{N}_{2}$.
We say that $\vaa{lts}_1\, \vaa{\sqsubseteq}_{lts} \,\vaa{lts}_2 $ iff $\vaa{M}_{0,1} \preccurlyeq_{lts} \vaa{M}_{0,2}$.
\end{definition}


The approximation order for abstract LTS is based on a 
simulation between abstract states. More in details, we say that 
$\vaa{M}_{2}$ {\em simulates} $\vaa{M}_{1}$ ($\vaa{M}_{1} \preccurlyeq_{lts} \vaa{M}_{2}$) whenever $\vaa{M}_{2}$ approximates $\vaa{M}_{1}$, and there exists a surjective function   $H_{t}:\mathsf{Ts}(\vaa{M_{1}}) \rightarrow
  \mathsf{Ts}(\vaa{M_{2}})$ between the transitions of $\vaa{M}_1$ and
$\vaa{M}_2$. In particular,
each move $\vaa{M}_{1} \aarrow{\Theta}{\vaa{\Delta}_{1}}{r} \vaa{N}_{1}$ 
has to be matched by a move  $\vaa{M}_{2} \aarrow{\Theta}{\vaa{\Delta}_{2}} {r} \vaa{N}_2$, related to the same label $\Theta$, and such that $\vaa{\Delta}_{1}\oi\vaa{\Delta}_{2}$, showing that the multiplicities are properly approximated.

The following theorem shows that the abstract LTS 
computed for an abstract state $\vaa{M}$ is a {\em sound approximation} of the LTS, for any
 $M$ represented by $\vaa{M}$. 
\begin{theorem}[Soundness]
\label{t:safeLTS}
 Let $E$ be an environment and    $\vaa{M}
\in \vaa{{\cal M}}$. For each $M' \in
\gamma(\vaa{M})$, we have 
$\alpha_{lts}(\lts((E, {M'})))\,\vaa{\sqsubseteq_{lts}}\,\vaa{\lts}((E, \vaa{M})).$
 \end{theorem}

Splitting {\em hybrid}  states by means of the $\bigtriangledown^{\sharp}$ operator, in order to distinguish terminated and  non-terminated states may,  in general, increase drastically  the number of  abstract states.
For example the abstract LTS, starting from 
the  state
$\vaa{M}_0=\{ ([1,2],X), $$([1,2],Y)\}$ w.r.t. to the environment $E$ 
of Example \ref{ese2}, would have  14 abstract states. 

It is worth noting,  however,  that for modeling probabilistic termination   we don't need to be too fine in distinguishing different non-terminated states.   
For this reason we can apply the following widening operator to each abstract transition step: we approximate the  new abstract state $\vaa{M}$, result of the application of the transition relation of Table~\ref{tab:atran},  with an abstract state $\vaa{M}_{1}$, if $\vaa{M\sqsubseteq_{I}}  \vaa{M}_{1}$ and $\vaa{M}_{1}$ was  already generated in a previous derivation step. 
This will reduce the number of new generated abstract states as it is shown in the next example. 

For  these reasons 
in the following we always assume the application of the previous widening operator. 

\begin{example} 
\label{e:eseLTS1}
Fig. \ref{fig306} shows the complete abstract LTS 
for the  abstract state
$\vaa{M}_0=\{ ([1,2],X), $$([1,2],Y)\}$ w.r.t. to the environment $E$ 
of Example \ref{ese2}, where

\[\begin{array}{llll}
\vaa{M_{1}}=\{([2,3],X), ([1,1],Y) \ \
\vaa{M_{2}}=\{([3,4],X), ([0,0],Y) \ \ 
\vaa{M_{3}}=\{([2,3],X), ([0,0],Y)\\
\vaa{M_{4}}=\{([0,0],X), ([2,3],Y)\ \
\vaa{M_{5}}=\{([1,1],X), ([2,3],Y) \ \
\vaa{M_{6}}=\{([0,0],X), ([3,4],Y) \ \
\end{array}\]

{\tiny
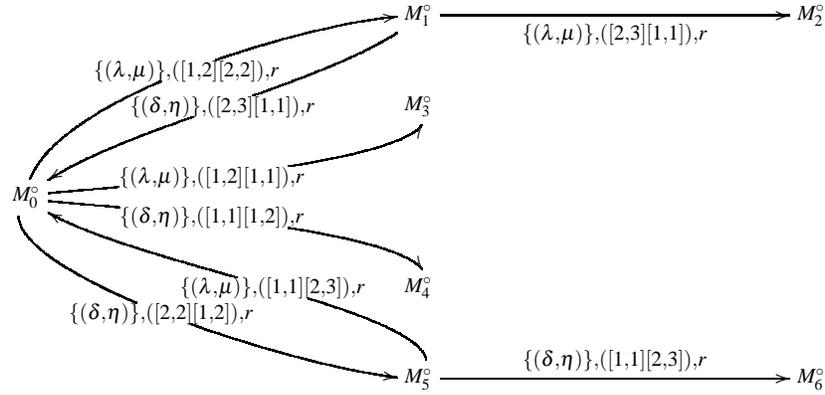
\begin{figure}

\begin{displaymath}
\def\objectstyle{\scriptstyle}
\def\labelstyle{\scriptstyle}
 \xymatrix@C+9pc@R-0.3pc{ 
 &
\vaa{M}_{1} 
\ar[r]_{\{(\lambda, \mu)\},([2,3][1,1]),r}
 \ar@(dl,ur)[ddl]|{\{(\delta, \eta)\}, ([2,3][1,1]),r} 
&   \vaa{M}_{2}\\
&\vaa{M}_{3}   \\
\vaa{M}_{0} \ar@(u,l)[uur]|{\{(\lambda, \mu)\},([1,2][2,2]) ,r} 
\ar@(l,dr)[ur]|{\{(\lambda, \mu)\},([1,2][1,1]) ,r} 
\ar@(l,ur)[dr]|{\{(\delta, \eta)\}, ([1,1][1,2]),r} 
\ar@(dl,l)[ddr]|{\{(\delta, \eta)\}, ([2,2][1,2]),r}     \\
&
\vaa{M}_{4}     \\
& 
\vaa{M}_{5} 
\ar[r]^{\{(\delta, \eta)\},([1,1][2,3]),r}
 \ar@(ur,dr)[uul]|{\{(\lambda, \mu)\}, ([1,1][2,3]),r} 
&   \vaa{M}_{6} }
\end{displaymath}
\caption{The abstract LTS}\label{fig306}
\end{figure}}

\end{example}

\section{Abstract Probabilistic Semantics}
\label{s:markova}

In standard Interval Markov Chains  \cite{SVA06,FLW06} 
transitions report  intervals of probabilities, representing a {\em lower}
and {\em upper} bound on the concrete probabilities, e.g.
a
set of possible distributions. Unfortunately, this information  is not adequate for our abstraction. 
Let us consider again the system, commented in Examples \ref{ese2} and 
\ref{e:eseLTS1}. As it is illustrated in the LTS of Fig.
\ref{fig306},
 the reachable states from $\vaa{M}_0$ are
$\vaa{M}_1$, $\vaa{M}_3$, $\vaa{M}_4$ and $\vaa{M}_5$ (see also  Fig. \ref{f:dafare} (c)).

In order to reason on the interval of probabilities we could safely assign to each transition leaving from $\vaa{M}_0$, it is useful to examine the set of concrete probability distributions, for each multiset $M_0$,
represented by $\vaa{M}_0$.
The DTMC corresponding to one of experiments represented by $\vaa{M}_0$ is described in Fig. \ref{fig300}; the other cases show analogous behaviors.
Actually, for each  $M_0$, there are two possible synchronizations between reagents $X$ and $Y$: one corresponding to the duplication of $X$ and the other
one corresponding to the duplication of $Y$. These  two alternative  moves always have the same probability.

Moreover,
each solution $M_0$, when there is a duplication of $X$, 
evolves into a solution, which is represented {\em either} by $\vaa{M}_1$ (where reagent $Y$ is still available) or by $\vaa{M}_3$ (where the concentration of $Y$ is $0$). 
Analogously, for the duplication of $Y$ and
the abstract states $\vaa{M}_4$ and $\vaa{M}_5$.
Thus, the abstract distributions representing the concrete distributions  are:
\[\begin{array}{l}
\rho_1(M_3)=1/2, \rho_1(M_1)=0, \rho_1(M_5)=1/2, \rho_1(M_4)=0,\\
\rho_2(M_3)=1/2, \rho_2(M_1)=0, \rho_2(M_5)=0, \rho_2(M_4)=1/2,\\
\rho_3(M_3)=0, \rho_3(M_1)=1/2, \rho_3(M_5)=1/2, \rho_3(M_4)=0,\\
\rho_4(M_3)=0,  \rho_4(M_1)=1/2, \rho_4(M_5)=0, \rho_4(M_4)=1/2.
\end{array}\]

It should be clear that
the {\em most precise} intervals of probabilities representing
the previous distributions, could be obtained
by considering  the {\em minimum}
and {\em maximum} probability, for each move. The  intervals we would obtain in this way,   are illustrated 
in Fig \ref{f:dafare} (a). This representation introduces a clear loss of information. For instance,
 the intervals include a  distribution such as
  $\rho(M_1)=1/2, \rho(M_3)=1/2, \rho(M_4)=0, 
\rho_4(M_5)=0$, 
which  does not correspond to any concrete behavior. Actually, 
states $\vaa{M}_1$ and $\vaa{M}_3$ are in {\em conflict}.

{\tiny
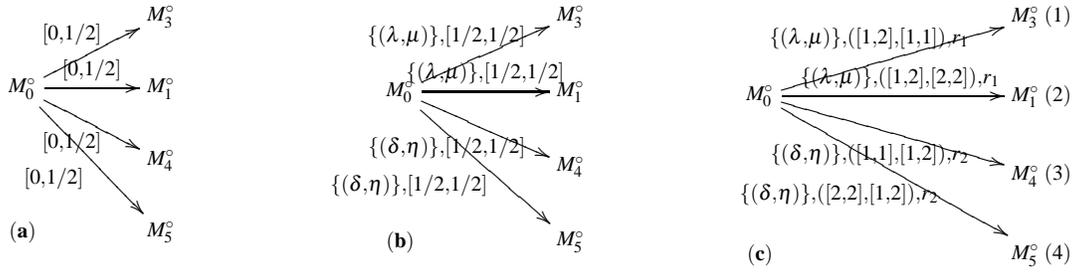
\begin{figure}
\[\begin{array}{lll}
\def\objectstyle{\scriptstyle}
\def\labelstyle{\scriptstyle}
 \xymatrix@C+1pc@R-0.9pc{ 
 &
 \vaa{M}_{3} 
  \\
   \vaa{M}_0 \ar[ur]^{[0,1/2]} 
\ar[r]^{[0,1/2]} 
\ar[dr]_{ [0,1/2]} 
\ar[ddr]_{ [0,1/2]}   &\vaa{M}_{1}   \\
&
\vaa{M}_{4}   \\
\bf{(a)}& 
 \vaa{M}_{5} }
   &\hspace*{1.5cm}
\def\objectstyle{\scriptstyle}
\def\labelstyle{\scriptstyle}
 \xymatrix@C+2pc@R-0.8pc{ 
 &
 \vaa{M}_{3} 
  \\
   \vaa{M}_0 \ar[ur]^{\{(\lambda, \mu)\},[1/2,1/2]} 
\ar[r]^{\{(\lambda, \mu)\},[1/2,1/2]} 
\ar[dr]_{\{(\delta, \eta)\}, [1/2,1/2]} 
\ar[ddr]_{\{(\delta, \eta)\}, [1/2,1/2]}   &\vaa{M}_{1}   \\
&
\vaa{M}_{4}   \\
\bf{(b)}& 
 \vaa{M}_{5} }
  & \hspace*{1.5cm}
\def\objectstyle{\scriptstyle}
\def\labelstyle{\scriptstyle}
 \xymatrix@C+5pc@R-0.7pc{ 
 &
 \vaa{M}_{3} \;(1)
  \\
   \vaa{M}_0 \ar[ur]^{\{(\lambda, \mu)\},([1,2],[1,1]),r_{1}} 
\ar[r]^{\{(\lambda, \mu)\},([1,2],[2,2]),r_{1}} 
\ar[dr]_{\{(\delta, \eta)\}, ([1,1],[1,2]),r_{2}} 
\ar[ddr]_{\{(\delta, \eta)\}, ([2,2],[1,2]),r_{2} }   &\vaa{M}_{1} \;(2)  \\
&
\vaa{M}_{4}\;(3)   \\
{\bf(c)}& 
 \vaa{M}_{5}\;(4) }
 \end{array}\]
\caption{The interval of probabilities and of multiplicities for $\vaa{M}_{0}$'s transitions.}\label{f:dafare}
\end{figure}
}

Since labels are suitably exploited in the abstract LTS in order to represent conflict,
we introduce a {\em generalization} of the original model,
called {\em Labeled Interval Markov Chains} (IMC). The model permits to more accurately represent the set of distributions represented by  intervals of probability by means of labels.\vspace*{0.3cm}\\
{\bf Labeled Interval Markov Chains.}
\begin{definition}[IMC]
  A IMC is a tuple
  $ (\abs S,\Pr^-, \Pr^+, \flabel,\abs M_0)$
  where
  \begin{enumerate}
  \item $\abs S \subseteq \abs{\mathcal{M}}$ is a countable set of abstract states and $\abs {M_0} \in \abs {S}$ is the \emph{initial state};
 \item 
    $\Pr^-, \Pr^+ \colon \abs S \to \SDistr(\abs S)$ are
    the \emph{lower} and \emph{upper bounds} on probabilities, such that
for each $\abs {M_1},\abs {M_2}
    \in \abs S$,
    $\Pr^-(\abs {M_1})(\abs {M_2}) \leq \Pr^+(\abs {M_1})(\abs {M_2})$;
  \item $\flabel: \abs S  \to ( \abs S \to \wp{(\widehat{{\cal L}}))}   $ is a {\em labeling function}.

\end{enumerate}
\end{definition}

In the following we use $\abs{{\cal IMC}}$ to denote the set of $IMC$.
As in the standard model, $\Pr^-(\abs {M_1})(\abs {M_2})$ and $\Pr^+(\abs {M_1})(\abs {M_2})$ define the {\em lower} and {\em upper} bound,  for the move from $\abs {M_1}$ to  $\abs {M_2}$, respectively. In addition, $\flabel(\abs {M_1})(\abs {M_2})$ reports the
set of labels corresponding to the move.
Intervals represent set of {\em admissible} distributions;
the notion of admissible distribution has to be slightly   adapted in order to handle the conflict between (sets of) labels.


\begin{definition}[Conflict of Labels]
\label{d:conflict}
Let $\alpha, \beta \in \wp ({\widehat{{\cal L}}})$ be sets of labels.
We say that $\alpha$ is in  {\em conflict} with $\beta$ iff
 there exists 
  $\vartheta \in {\widehat{\cal L} }$
 such that $\alpha=\{\vartheta\}=\beta$. 
\end{definition}
The notion of conflict between labels obviously induces a corresponding notion of conflict between states.  Let   $ (\abs S,\Pr^-, \Pr^+, \flabel,\abs M_0)$ be an IMC and $\abs M \in \abs{S}$.
We say that $\vaa{NS} \subseteq \abs{S}$ is a set of  \emph{ no-conflict states} w.r.t. $\vaa{M}$ iff it is  maximal   and, for each  $\abs{M}_{1},\vaa{M}_2 \in \vaa{NS}$, there is no conflict between $\flabel(\vaa{M})(\vaa{M}_1)$ 
and $\flabel(\vaa{M})concept\vaa{M}_2)$.

 


\begin{definition}[Admissible Distribution] 
\label{d:admissible}
  Let   $\abs{mc}=(\abs S,\Pr^-, \Pr^+, \flabel,\abs M_0)$ be an IMC and let $\abs M \in \abs S$. 
We say that  a distribution $\rho \in \Distr(\abs S)$ is {\em admissible} for $\abs M$ iff
there exists a set of  \emph{ no-conflict states} $\vaa{NS}$ such that,
for each $\abs{M}_1 \in \abs S$:  if $\abs {M}_1\in \vaa{NS}$, then $\Prm(\abs{M})(\abs {M}_1) \leq \rho(\abs {M}_1) \leq \Prp(\abs M)(\abs {M}_1)$; $\rho(\abs {M}_1) =0$, otherwise.
 We use $\ADistr_{\abs{mc}}(\abs M)$ for the set of admissible distributions for $\abs M$.

\end{definition}

Intuitively, an admissible  distribution $\rho$  corresponds to a set of no-conflict states $\vaa{NS}$, and reports a value included in the interval, 
for each state of $\vaa{NS}$, and zero otherwise. 
As an example, the IMC illustrated in Fig \ref{f:dafare} (b) reports four non-conflict set of states w.r.t. $\vaa{M}_0$: (1) $\{\vaa{M}_3,\vaa{M}_4\}$, (2) $\{\vaa{M}_3,\vaa{M}_5\}$; (3) $\{\vaa{M}_1,\vaa{M}_4\}$ and (4)
$\{\vaa{M}_1,\vaa{M}_5\}$. As a consequence,  the admissible distributions, corresponding
to (1)-(4) are exactly the distributions $\rho_1-\rho_4$, discussed at the beginning of the Section.
This shows that the IMC of Fig. \ref{f:dafare} (b) is a sound (and very precise) approximation of the probabilistic
semantics, for each multiset represented by $\vaa{M}_0$. 

Once defined admissible distributions the concept of scheduler follows the same guidelines of \cite{GL08}.  
The notion of path  and cylinder for IMC are  analogous
to that presented for DTMC.

\begin{definition}[Scheduler]
   Let   $\abs{mc}= (\abs S,\Pr^-, \Pr^+, \flabel ,\abs M_0)$ be an IMC, a \emph{scheduler}  is a function $A
  \colon $ $\FPaths(\abs S) \to \Distr(\abs S)$ such that $A(\abs\pi) \in
  \ADistr_{\abs{mc}}(\abs\pi[|\abs\pi|])$ for any abstract path
  $\abs \pi \in \FPaths(\abs S)$.  
We use $\Adv(\abs{mc})$ to denote the set of schedulers.
  
\end{definition}

Given a scheduler  a probability space
over paths can be defined analogously as for DTMC. In the following,
 $\Pr_{\abs M}^{\Pi} \in \Adv(\abs{mc})$  stands for the probability starting from  $\abs M$ w.r.t. the scheduler $\Pi \in Adv(\abs{mc})$.

An IMC gives both
 {\em under} and  {\em over} approximations of the probability
of reachability properties, that can be computed by considering
the {\em worst} and {\em best} probabilities w.r.t. all the schedulers. For approximating
{probabilistic termination}, we have to define terminated  abstract states.
A state $\abs{M}\in \abs S$ of a IMC $\abs{mc}=(\abs S,\Pr^-, \Pr^+, \flabel,,\abs M_0)$ 
is $\exists$-\emph{terminated}  iff  $\Pr^+(\abs M)(\abs M)=1$, and 
is $\forall$-\emph{terminated}  iff  $\Pr^-(\abs M)(\abs M)=1$.

\begin{definition}[Probabilistic Termination]
Let   $\abs{mc}=(\abs S,\Pr^-, \Pr^+, \flabel,,\abs M_0)$ be an IMC.  The \emph{lower} and \emph{upper bound} of probabilistic termination, starting from
  $\abs M \in \abs S$, are 
  \[\begin{array}{rl}
    \Reach^-_{\abs{mc}}(\abs M) &= \inf_{\Pi \in \Adv(\vaa{mc})} \Pr_{\abs M}^\Pi (\{ \abs\pi \in
    C(\abs M) \mid \abs\pi[i]  \text{ is $\forall$-terminated for some $i \geq 0$}
    \}) \\
    \Reach^+_{\abs{mc}}(\abs M) &= \sup_{\Pi \in \Adv(\vaa{mc})} \Pr_{\abs M}^\Pi (\{ \abs\pi \in
    C(\abs M) \mid \abs\pi[i] \text{ is $\exists$-terminated  for some $i \geq 0$}
    \})
  \end{array}\]

\end{definition}

Finally, we observe that 
the problem of   model checking the  IMC can be reduced, as in the case of Markov Interval Chains,   to the verification
of a  Markov Decision Process (MDP), by considering the  so called feasible solutions.
The complexity of this   reduction is comparable to  the one for a standard  Markov Interval Chains with the same number of states. 
Analogously,  more efficient iterative algorithms which construct a basic  feasible solution  \emph{ on-the-fly}  can also be used to model check our  IMC   (see \cite{SVA06,FLW06}).  
  \vspace*{0.3cm}\\
{\bf Soundness and precision of approximations.}\vspace*{0.3cm}\\
We introduce  a notion of {\em best abstraction} of a  DTMC based on an  approximation order on IMC. Here, for a lack of space, we give just an intuitive definition of such an  order. The reader can refer to \cite{GL08} for the formal definition.

\begin{definition}[Best Abstraction]
\label{d:bam}
 We define $\alpha_{MC}: {\cal MC} \rightarrow \abs{\cal IMC}$ such that
$\alpha_{MC} ((S, \Pr,\flabel, M_0)) = (\{ \alpha(M)\}_{M\in S},  {\Pr_{\alpha}}^-, {\Pr_{\alpha}}^+,\flabel, \alpha(M_0))$,  where $ {\Pr_{\alpha}}^-(\alpha({{M_1}}),  \alpha({{M_2}})) ={\Pr_{\alpha}}^+(\alpha({{M_1}}),  \alpha({{M_2}}))  =\Pr(M_1)(M_2)$.
\end{definition}


  


The order on IMC is based on  a sort of probabilistic simulation. Intuitively,  $\abs {M_2}$ simulates $\abs {M_1}$ ($\abs {M_1}\preccurlyeq_{mc}\abs {M_2}$) whenever: (i) 
$\abs {M_2}$ approximates $\abs {M_1}$: (ii)
 each distribution of $\abs {M_1}$ is matched by a corresponding distribution of $\abs {M_2}$, where the probabilities of the target
states are eventually summed up.

This simulation provides sufficient conditions for the
preservation of extremum probabilities, as stated by the following theorem.

\begin{theorem}[Soundness of the order]
\label{teo:sorde}
Let $\abs{mc}_i= (\abs {S_i},\Pr_i^-, \Pr_i^+,\flabel_{i}, \abs M_{0,i})$
be two IMC and let $\abs {M_i} \in \abs {S_i}$,
for $i \in \{1,2\}$.
  If  $\abs {M_1} \preccurlyeq_{mc} \abs {M_2}$, then   $
    \Reach_{\abs{mc}_2} ^-(\abs {M_2}) \leq \Reach_{\abs{mc}_1}^-(\abs {M_1}) \leq \Reach_{\abs{mc}_1}^+(\abs {M_1}) \leq \Reach_{\abs{mc}_2}^+(\abs {M_2})$.
\end{theorem}

\section{Derivation of IMC}
\label{s:aps}
We define a systematic method for deriving an IMC from an abstract LTS.
Obviously, the crucial part of the translation consists of the calculation of {\em intervals of probabilities}
from the information reported on abstract transitions labels. 
The approach, proposed in \cite{GL08}, suggests a methodology similar
to the one applied in the concrete case, based on the calculation of {\em 
abstract rates}, e.g. intervals of rates.

The idea is to derive from abstract transition labels 
 the interval of rates $\vaa{\rate}(\vaa{t})$ corresponding to any abstract transition $\vaa{t}$.
Then, by "summing up" the abstract rates $\vaa{\rate}(\vaa{t})$ of all transitions  $\vaa{t} \in \mathsf{Ts}
(\vaa{M}_1,\vaa{M}_2)$, we can obtain the abstract rate 
 $\vaa{\Rat}(\vaa{M}_1,\vaa{M}_2)$ 
  for the complete move from $\vaa{M}_1$
to $\vaa{{M}}_2$. Analogously, we can also obtain the abstract exit rate $ \vaa{\E}(\vaa{M}_1)$ corresponding to all the moves from $\vaa{M}_1$. 
Finally, both  {\em lower}  and {\em upper}  bounds of the  probability of moving from 
 $\vaa{M}_1$ to $\vaa{{M}}_2$
can  easily be computed by {\em minimizing} and {\em maximizing} the solution of  
$\vaa{ \Rat}(\vaa{M}_1,\vaa{{M}}_2) \vaa{/} \vaa{\E}(\vaa{M}_1)$, resp..

However, the refined abstract LTS semantics presents a relevant difference: the labels represent a notion of {\em conflict} between abstract transitions.
As an example,   Fig. \ref{f:dafare} (c) reports the abstract transitions (see also Example \ref{e:eseLTS1} and Fig. \ref{fig306}) for the   abstract state $\vaa{M}_0$. 
Notice that just four combinations of transitions are possible: (a) (1) and (3);
(b) (1) and (4); (c) 
(2) and (3);
(d) (2) and (4). 
It should be clear that each combination $i \in \{(a)-(d)\}$ leads to a
 different  {\em abstract exit rate} for $\vaa{M}_0$, $\vaa{\E}_i(\vaa{M}_1)$. As a consequence, in order to generalize
the approach of \cite{GL08},
we could {\em minimize} and {\em maximize} the solution of   $\vaa{ \Rat}_i(\vaa{M}_1,\vaa{{M}}_2) \vaa{/} \vaa{\E}_i(\vaa{M}_1)$,
for each  combination $i \in \{(a)-(d)\}$,  
 resp..

It should be clear that this naive generalization of the approach would be very computationally expensive. Therefore, we propose a more efficient approximated calculation.
The idea is to compute a different {\em exit rate} $ \vaa{\E}_{\vaa{M}_2}(\vaa{M}_1)$ for $\vaa{M}_1$,
w.r.t. each $\vaa{M}_2$, reporting the {abstract  rate} of \emph{all}  transitions which may appear in parallel with a 
 transition  of  $\mathsf{Ts}
(\vaa{M}_1,\vaa{M}_2)$. This represents obviously an approximation 
    of the exit rates that we would obtain by considering all combinations involving a
   transition  of $\mathsf{Ts}
(\vaa{M}_1,\vaa{M}_2)$.

In the style of \cite{GL08}, the abstract rates (intervals of rates) are represented by {\em symbolic expressions} on reagent variables, such as
$(e,c)$, where:
(i) $e \in {\cal Z}$ is an {\em expression} over  variables  $\cal X$; (ii)
$c \in {\cal C}$ is a set of {\em membership constraints} of the 
form $X \in I$ \footnote{We require that,  $\forall X\in Vars(e)$,  there exists
exactly one constraint $X \in I$ in $c$.}. This approach permits to more accurately exploit the information recorded by abstract transition labels.
Moreover, for $op \in\{+,/ \}$ we use: (a) 
$(e_{1},c_{1})\vaa{op}( e_{2},c_{2})=(e_{1}\; op \;e_{2},  c_1 \cup c_2)$; (b) $
(e,c_{1}) \vaa{\cup} ( e,c_{2})=(e,  c_1 \cup c_2),\mbox{ where } 
c_1 \cup c_2=\bigcup_{X\in {\cal X}} ( X \in {\bigsqcup_{I}}_{(X\in I)\in c_{i}, i\in\{1,2\}}I 
)).$

The {\em abstract rate} of  a transition
$\vaa{t}=\vaa{M}_1 \aarrow {\Theta} {\vaa{\Delta}}r \vaa{M}_{2} $ can be defined
as follows

\[\vaa{\rate}(\vaa{t})= \left \{ \begin{array}{llll} (X \cdot r , \{X\in I\})&
\Theta= \lambda, \lambda \in {\cal L}(E.X) , \vaa{\Delta}=I,\\
(X \cdot (X \widehat{-}1) \cdot r,\{X\in I\} )&\Theta=(\lambda,\mu), \vaa{\Delta}= (I,I), \lambda,\mu \in {\cal L}(E.X),\\
(X \cdot Y \cdot r, \{X \in I_{1}, Y \in I_2\}) &
\Theta=(\lambda,\mu), \vaa{\Delta}= (I_{1},I_{2}),  
\lambda \in {\cal L}(E.X),
 \mu \in {\cal L}(E.Y), X \neq Y. 
 \end{array}\right.
\]

Then, we define $\vaa{\E}_{\vaa{M}_2}(\vaa{M}_1)$ and $\vaa{\Rat}(\vaa{M}_1,\vaa{M_2})$, where $\vaa{Ts}
\subseteq \mathsf{Ts}(\vaa{M}_1 )$, 
\[\begin{array}{ll}
\vaa{\E}_{\vaa{M}_2}(\vaa{M}_1)= 
\vaa{\sum}_{ (e,c) \in \rate({\mathsf{Ts}}_{\setminus \vaa{M}_2}(\vaa{M}_1)\cup
\mathsf{Ts}
(\vaa{M}_1,\vaa{M}_2) )}
(e,c)
&\;\;
\vaa{\Rat}(\vaa{M}_1,\vaa{M_2})=\vaa{\sum}_{  \vaa{t}\in \mathsf{Ts}(\vaa{M}_1,\vaa{{M}}_2)} \widehat{\vaa{\rate}}(\vaa{t})
\end{array}\]
\[\begin{array}{l}
\widehat{\vaa{\rate}}(\vaa{t})= \left \{\begin{array}{ll}
(e,c\cup \{\;X\in[0,0]\;|\;X\in Vars(e)\})  & \mbox{if } \vaa{\rate}(\vaa{t})=(e,c)\mbox{ and }
\lab(\vaa{t}) \in \lab ({\mathsf{Ts}}_{\setminus \vaa{M}_2}(\vaa{M}_1)),\\
\vaa{\rate}(\vaa{t}) & \mbox{otherwise.}
\end{array} \right.\\
\\
\rate( \vaa{Ts})
=\{r_{\Theta} \mid \Theta \in \widehat{{\cal L}}, r_{\Theta}= \vaa{\bigcup}_{\{\vaa{t} \in  
 \vaa{Ts}, \lab(\vaa{t})=\Theta \}} \rate(\vaa{t})\}\\
 \\
 {\mathsf{Ts}}_{\setminus \vaa{M}_2}(\vaa{M}_1)=\{\vaa{t}\in\mathsf{Ts}(\vaa{M}_1)
|\target(\vaa{t}) \neq \vaa{M}_2,  \lab(\vaa{t}) \mbox{ not in conflict
with } 
 \lab(\mathsf{Ts}(\vaa{M}_1,\vaa{M}_2 ))\}
\end{array}\]

Here, $\mathsf{Ts}_{\setminus \vaa{M}_2}(\vaa{M}_1) \subseteq \mathsf{Ts}(\vaa{M}_1) $ reports the transitions which may appear in parallel with a transition of $\mathsf{Ts}
(\vaa{M}_1,\vaa{M}_2)$. 
In the calculation of $\vaa{\E}_{\vaa{M}_1}(\vaa{M}_2)$ the abstract  rates of
 transitions  with   the same label are merged (namely approximated)  by taking the union of the membership constraints.

Finally, both  {\em lower}  and {\em upper}  bounds of the  probability of moving from 
 $\vaa{M}_1$ to $\vaa{{M}}_2$
can be derived by minimizing and maximizing the solution of  
$\vaa{ \Rat}(\vaa{M}_1,\vaa{{M}}_2) \vaa{/} \vaa{\E}_{\vaa{M}_2}(\vaa{M}_1)$, resp..
This reasoning has to be properly combined with two special cases
when $max(\vaa{\E}_{\vaa{M}_{2}}(\vaa{M}_{1})) =0$ or $
min(\vaa{\E}_{\vaa{M}_{2}}(\vaa{M}_{1})) =0$.


\begin{definition}
\label{d:anormal}
The {\em abstract probabilistic translation} function $\vaa{\N }:\vaa{{\cal LTS}} \rightarrow  \vaa{{\cal IMC}}$
such that \\
    $\vaa{\N}((\vaa{S},\vaa{\rightarrow},\vaa{M_{0}},E)) =  (\abs S,\Pr^-, \Pr^+, \flabel,\abs M_0)$, and  
    $\Pr^-,$ $\Pr^+ \colon \abs S \to \SDistr(\abs S)$ are
    the \emph{lower} and \emph{upper} probability functions,
    such that for each $\vaa{M}_{1}\in \abs S$: 
    \begin{description}
    \item[a)]
for each 
 $ \vaa{M}_{2} \in \vaa{S},$ such that $ max(\vaa{\E}_{\vaa{M}_{2}}(\vaa{M}_{1})) >0$,
 if 
$min(\vaa{\Rat}(\vaa{M}_{1},\vaa{M}_{2}))=0,$ then also
$\Pr^{-}(\vaa{M}_{1})(\vaa{M}_{2})=0,$  
otherwise, 
$\Pr^{-}(\vaa{M}_{1})(\vaa{M}_{2})=min(\vaa{ \Rat}(\vaa{M}_{1},\vaa{M}_{2}) \vaa{/} \vaa{\E}_{\vaa{M}_{2}}(\vaa{M}_{1}))$. Analogously, the  $\Pr^{+}$ function  is obtained by substituting in the previous definition,  the  $min$ function with the $max$ function;
    \item[b)]
if, for each $\vaa{M}_{2} \in \vaa{S}$, $ max(\vaa{\E}_{\vaa{M}_{2}}(\vaa{M}_{1})) =0$, then $\Pr^{+}=\Pr^{-}$, 
$\Pr^{+}(\vaa{M}_{1})(\vaa{M}_{1})= 1$, and  $\forall \vaa{M}_{2}\not= \vaa{M}_{1}$, \\
$\Pr^{+}( \vaa{M}_{1}),( \vaa{M}_{2})=
0 $;
    \item[c)]
if,  $\exists \vaa{M}_{2} \in \vaa{S}$, such that $ max(\vaa{\E}_{\vaa{M}_{2}}(\vaa{M}_{1})) >0$
and  $min(\vaa{\E}_{\vaa{M}_{2}}(\vaa{M}_{1}) )=0$ then 
$\Pr^{+}(\vaa{M}_{1})(\vaa{M}_{1})= 1$, and 
$\Pr^{-}( \vaa{M}_{1}), (\vaa{M}_{1})=
0 $.
\end{description}
$\flabel: \abs S  \to (\abs S \to  \wp ({\widehat{{\cal L}}}))$  is a labeling function defined as 
 $ \forall \vaa{M}_{1},\vaa{M}_{2} \in \vaa{S}$, $\flabel(\vaa{M}_{1},\vaa{M}_{2})=\lab(\{\vaa{t} \in \mathsf{Ts}(\vaa{M}_{1},\vaa{M}_{2}) \mid max(\vaa{\rate}(\vaa{t}))> 0\}).$
\end{definition}


 The following theorems state the soundness of our approach.

\begin{theorem}
\label{teo:soundn1}
Let $\abs{lts}_i=(\abs{S_i},\abs{\rightarrow_i}, \abs{M_{0,i}},E)$
be two abstract LTS. If 
$\abs{lts}_1\, \vaa{\sqsubseteq}_{lts}\, \abs{lts}_2$, then 
$\abs{\N}(\abs{lts}_1)\, \vaa{\sqsubseteq}_{mc}\, \abs{\N}(\abs{lts}_2)$.

\end{theorem}

\begin{theorem}
\label{teo:soundn2}
Let  $E$ be an environment and  $M_{0} \in {\cal M} $ be a multiset.

We have $\alpha_{MC}(\N(\lts((E, M_0)))) \,\vaa{\sqsubseteq}_{mc}\,
\abs{\N}(\alpha_{lts}(\lts((E, M_0)))).$

\end{theorem}

\begin{example}
\label{e:eseastr1}
Fig. \ref{fig301} describes the IMC, obtained from the abstract LTS
of Fig. \ref{fig306}, for the  abstract state 
$\vaa{M}_0=\{ ([1,2],X), $$([1,2],Y)\}$ (see also Examples 
  \ref{ese2} and 
\ref{e:eseLTS1}).

{\tiny
\begin{figure}
\begin{displaymath}
\def\objectstyle{\scriptstyle}
\def\labelstyle{\scriptstyle}
 \xymatrix@C+5pc@R-0.4pc{ 
 &
{ \vaa{M}_{1}} 
\ar[r]^{\{(\lambda, \mu)\},[1/2,1/2]}
 \ar@(dl,ur)[ddl]|{\{(\delta, \eta)\}, [1/2,1/2]} 
&    \vaa{M}_{2}
\ar@(dr,ur)[]|{\hspace* {0.4cm}\emptyset,[1,1]}  
  \\
&
 \vaa{M}_{3}\ar@(dr,ur)[]|{\hspace* {0.4cm}\emptyset,[1,1]}   \\
{ \vaa{M}_0} \ar@(u,l)[uur]|{\{(\lambda, \mu)\},[1/2,1/2]} 
\ar@(l,dr)[ur]|{\{(\lambda, \mu)\},[1/2,1/2]} 
\ar@(l,ur)[dr]|{\{(\delta, \eta)\}, [1/2,1/2]} 
\ar@(dl,l)[ddr]|{\{(\delta, \eta)\}, [1/2,1/2]}     \\
&
\vaa{M}_{4} \ar@(dr,ur)[]|{\hspace* {0.4cm}\emptyset,[1,1]}    
   \\
& 
{ \vaa{M}_{5} }
\ar[r]_{\{(\delta, \eta)\},[1/2,1/2]}
\ar@(ur,dr)[uul]|{\{(\lambda, \mu)\}, [1/2,1/2]} 
 &  {\vaa{M}_{6}} 
\ar@(dr,ur)[]|{\hspace* {0.3cm}\emptyset,[1,1]}} 
 \end{displaymath}
\caption{The IMC}\label{fig301}
\end{figure}}
Note that the result is very precise. For $\vaa{M}_0$ we derive precisely the approximation,
discussed in Fig. \ref{f:dafare} (b); namely, four admissible  distributions corresponding to the combinations
of labels not in conflict. For the other states there is exactly
one admissible distribution. In particular, $\vaa{M}_2$, 
$\vaa{M}_3$, $\vaa{M}_4$ and $\vaa{M}_6$ are $\forall$-terminated.
By computing {\em lower} and {\em upper} bounds for 
 probabilistic termination, from $\vaa{M}_0$, we obtain
exactly one in both cases.
For  the  maximum, it is enough  to choose the admissible  distributions which reach terminated states as soon as possible. This is obviously represented
by the distribution for $\vaa{M}_0$, reporting probability $ 1/2$
to move  in $\vaa{M}_3$ and $\vaa{M}_4$.
By contrast, for  the  minimum, it is enough  to choose the admissible  distributions which do not reach terminated states,
every time this is possible. This is obviously represented
by the choice of the distribution for $\vaa{M}_0$, reporting probability $ 1/2$
to move  in $\vaa{M}_1$ and $\vaa{M}_5$. Thus, we obtain a DTMC, and the reasoning is
similar to that discussed in Example \ref{ese2}.

This proves  that each experiment, represented by $\vaa{M}_0$, leads to a terminated state with probability one, e.g.
universally terminates. Note that here we have examined a very small example
for sake of simplicity; however, it should be clear that the result
could be generalized to any concentration of reagents $X$ and $Y$.\hfill $\Box$
\end{example}

\section{Conclusions}
\label{s:conclu}
The methodology proposed in this paper is substantially different from most of the approaches, proposed in literature \cite{DJJL01,FLW06,Hu05,Mon05,KNP06,FLW06}, in order to abstract probabilistic models, based on abstract interpretation or
partitioning of the concrete state space.
Actually, our goal is to represent by means of the IMC of an abstract system a set of concrete systems, each corresponding to a different DTMC.
In this setting it is therefore essential to develop
an effective method (even for infinite state systems) for computing the abstract probabilistic model, directly from the abstract LTS.
The main contribution of the approach consists  in the calculation of the intervals of probabilities
from the information reported on abstract transition labels, without
building  all the concrete distributions. We have also shown that the technique of \cite{GL08} can be successfully generalized to the refined abstract LTS,
by finding out a good trade-off between precision and 
complexity. 
For this reason, a probabilistic  model
such as a Markov Decision Process is not adequate.

An advantage of our framework is that other kinds of uncertainties
of biological systems could be handled in a similar way.
For example, the approach could be easily adapted in order to model
(even infinite) sets of concrete systems with different values for the rates.
Another advantage of our framework, based on abstract interpretation, is
that new
analyses  could be easily designed by introducing  new abstract LTS semantics.
For example,  we would like to investigate the application of more precise numerical domains able to model also relational information,
such as the domain of convex polyhedra. 
We leave to the future work the extension of the framework
 to  the full calculus
with communication \cite{PC07} as well as the extension to Continuous-Time Markov Chains.

\bibliographystyle{eptcs}

\begin{thebibliography}{1}






\bibitem{BDNN01} 
C. Bodei, P.Degano, F.Nielson and H.Riis Nielson.
{\em Static Analysis for the Pi-Calculus with Applications to Security.}
\newblock Information and Computation, 168: 68-92, 2001.

 



\bibitem{Ca04}
L.~Cardelli.
\newblock {\em Brane Calculi.}
\newblock Proc. of  {CMSB} '04,  LNCS 3082, 257--278, 2004.

\bibitem{Ca07}
L.~Cardelli.
\newblock {\em On Process Rate Semantics.}
\newblock  Theoretical Computer Science, 391 190--215, 2008.

\bibitem{Ca09}
L.~Cardelli.
\newblock {\em Algorithmic Bioprocesses.}
In A.Condon, D.Harel, J.N.Kok, A.Salomaa, E.Winfree (Eds.), Springer, 2009






\bibitem{GL08}
A. Coletta and R.Gori and F. Levi.
\newblock {\em Approximating probabilistic behaviours of biological systems using abstract interpretation.}
\newblock Proc. of FBTC '08,  ENTCS 229 (1), 165--182, 2009.

\bibitem{CousotC76}
P.~Cousot and R.~Cousot.
\newblock {\em Static Determination of Dynamic Properties of Programs}.
\newblock \newblock {Proc. of POPL'76} ,
106--130, 1976.





\bibitem{CousotC77}
P.~Cousot and R.~Cousot.
\newblock {\em Abstract {I}nterpretation: {A} {U}nified {L}attice {M}odel for
  {S}tatic {A}nalysis of {P}rograms by {C}onstruction or {A}pproximation of
  {F}ixpoints}.
\newblock {Proc. of POPL'77},
   238--252, 1977.
   
   
\bibitem{CousotC79}
P.~Cousot and R.~Cousot.
\newblock {\em Systematic {D}esign of {P}rogram {A}nalysis {F}rameworks.}
\newblock Proc. of POPL'79 ,
 269--282, 1979.

\bibitem{CousotC91}
P.~Cousot and R.~Cousot.
\newblock {\em Comparing the {G}alois {C}onnection and {W}idening/{N}arrowing
  {A}pproaches to {A}bstract {I}nterpretation.}
\newblock  { Proc. of PLILP'92}, LNCS 631,
   269--295, 1992.

\bibitem{DGG94}
 D. Dams, R. Gerth  and O. Grumberg.
 \newblock {\em Abstract Interpretation of Reactive Systems.}
 \newblock  TOPLAS, 19(2), 253-291, 1997.

\bibitem{DJJL01}
 P. D'Argenio, B. Jeannet, H. Jensen and K. Larsen.
 \newblock {\em Reachability Analysis of Probabilistic Systems by Successive Refinements.}
 \newblock  { Proc. of PAPM-PROMIV'01}, LNCS 2165,
   39--56, 2001.

\bibitem{DJJL02}
 P. D'Argenio, B. Jeannet, H. Jensen and K. Larsen.
 \newblock {\em Reduction and Refinement Strategies for Probabilistic Analysis.}
 \newblock  { Proc. of PAPM-PROMIV'02}, LNCS 2399,
   57--76, 2002.

\bibitem{FLW06}
H. Fecher, M. Leucker and V. Wolf.
\newblock {\em Don't Know in Probabilistic Systems.}
\newblock Proc. of   SPIN'06, LNCS 3925, 71--88, 
       2006.


\bibitem{Fe01}
J. Feret.
\newblock {\em Abstract Interpretation-Based Static Analysis of
Mobile Ambients.}
\newblock Proc. of   SAS'01, LNCS
        2126,  412-430, Springer Verlag,
       2001.






\bibitem{GL05}
R.Gori and F. Levi.
\newblock {\em A new occurrence Counting analysis for BioAmbients.}
\newblock Proc. of APLAS '05,  LNCS 3780, 381--400, 2005.

\bibitem{GL06}
R.Gori and F. Levi.
\newblock {\em An Analysis for proving Temporal Properties of Biological Systems.}
\newblock Proc. of APLAS '06, LNCS 4279, 234--252, 2006.





\bibitem{HJ94}
H.Hansson and B. Jonsson.
\newblock{\em A  Logic for Reasoning about Time and Probability.}
\newblock Formal Aspects of Computing,  6(5), 512--535, 1994.



\bibitem{HKN06}
A. Hinton, M. Kwiatkowska, G. Norma and D. Parker.
\newblock{\em PRISM: a tool for automatic verification of probabilistic systems.}
\newblock Proc. of TACAS'06, LNCS 3920,  441-444,
Springer-Verlag, 2006.





\bibitem{Hu05}
M. Huth. 
\newblock{\em  On finite-state approximants for probabilistic computation tree logic.}
\newblock Theoretical Computer Science, 346(1),  113--134, 2005.


\bibitem{KSK76}
 J.G. Kemeny, J.L. Snell and A.W. Knapp.
\newblock{\em Denumerable Markov Chains.} 
\newblock Springer, 1976.  

\bibitem{Kw03}
 M. Kwiatkowska.
\newblock{\em Model checking for probability and time: from theory to practice.} 
\newblock Proc. of LICS' 03, 351--360, 2003.  

\bibitem{KNP06}
 M. Kwiatkowska, G. Norman and D. Parker.
\newblock{\em Game-based Abstraction for Markov Decision Processes.} 
\newblock Proc. of QEST'06, 157--166, 2006.  


\bibitem{LM04}
F. Levi and S. Maffeis.
\newblock {\em On Abstract Interpretation of Mobile Ambients.}
\newblock Information and Computation 188, 179--240,
2004.

%

%

%




\bibitem{Mon05}
D. Monniaux.
\newblock {\em Abstract interpretation of programs as Markov Decision Processes.}
\newblock Science of Computer Programming, 58(1-2), 179--205, 2005.



\bibitem{NNH02}
F. Nielson, H.R. Nielson,  R.R. Hansen.
\newblock{\em Validating firewalls using flow logics.}
\newblock Theoretical Computer Science, 283(2), 381-418, 2002.



\bibitem{NNP04}
F. Nielson, H.R. Nielson and  H. Pilegaard.
\newblock{\em Spatial Analysis of BioAmbients.}
\newblock Proc. of SAS'04, LNCS 3148, pp. 69--83,
Springer-Verlag, 2004.




\bibitem{PC04}
A. Phillips and L. Cardelli.
\newblock{\em A Correct Abstract Machine for the Stochastic Pi-calculus.}
\newblock Proc. of  {BioCONCUR} '04,  ENTCS, 2004.

\bibitem{PC07}
A. Phillips and L. Cardelli.
\newblock{\em Efficient, Correct Simulation of Biological Processes in the Stochastic Pi-calculus.}
\newblock Proc. of  {CMSB} '07, LNCS 4695, 184--199, 2007.


\bibitem{Pr05}
C.Priami. 
\newblock{\em Stochastic $\pi$-calculus.}
 \newblock The Computer Journal, 38, 578--589,1995.


\bibitem{PQ05}
C.Priami and P. Quaglia. 
\newblock{\em Beta binders for biological interactions.}
 \newblock Proc. of CMSB'04, LNCS 3082,20--33,2005.
\bibitem{PRSS01} 
C. Priami, A. Regev,  W. Silverman and E. Shapiro.
\newblock{\em Application of a stochastic name-passing calculus to representation and simulation of molecular processes.}
\newblock {Information Processing Letters, 80 (1), 25--31, 2001.}


\bibitem{RPSCS04} 
A. Regev, E. M. Panina, W. Silverman, L. Cardelli and E. Shapiro.
\newblock{\em BioAmbients: an Abstraction for Biological Compartments.}
\newblock {Theoretical Computer Science, 325, 141--167, 2004.}

\bibitem{RSS01} 
 A. Regev,  W. Silverman and E. Shapiro.
\newblock{\em Representation and Simulation of Biochemical Processes using the pi-calculus process algebra.}
\newblock {Proc. of the Pacific Symposium on Biocomputing 2001, 6, 459--470, 2001. }


\bibitem{SVA06}
K. Sen, M. Viswanathan and G. Agha.
\newblock{\em Model Checking Markov Chains in the Presence of Uncernainties.}
\newblock Proc. of TACAS'06, LNCS 3920, 394-410, 2006.


\bibitem{ZC08}
G. Zavattaro and L. Cardelli.
\newblock{\em Termination Problems in Chemical Kinetics.}
\newblock Proc. of CONCUR'08, LNCS 5201, 477-491, 2008.









\end{thebibliography}

\end{document}